\newcommand{\BR}{{\cal B}}

\newcommand{\piz}{\pi^0}

\newcommand{\ks}{K_S^0}

\newcommand{\psp}{\psi(2S)}
\newcommand{\jpsi}{J/\psi}
\newcommand{\EE}{e^+e^-}

\newcommand{\pp}{\pi^+\pi^-}

\newcommand{\kk}{K^+K^-}

\newcommand{\beq}{\begin{equation}}
\newcommand{\eeq}{\end{equation}}
\newcommand{\beqy}{\begin{eqnarray}}
\newcommand{\eeqy}{\end{eqnarray}}
\newcommand{\bitm}{\begin{itemize}}
\newcommand{\eitm}{\end{itemize}}

\documentclass{ws-ijmpcs}

\begin{document}

\markboth{Chengping Shen}
{Test of QCD at large $Q^2$ with exclusive hadronic processes}

%%%%%%%%%%%%%%%%%%%%% Publisher's Area please ignore %%%%%%%%%%%%%%%
%
\catchline{}{}{}{}{}
%
%%%%%%%%%%%%%%%%%%%%%%%%%%%%%%%%%%%%%%%%%%%%%%%%%%%%%%%%%%%%%%%%%%%%

\title{Test of QCD at large $Q^2$ with exclusive hadronic processes}

\author{Chengping Shen for the Belle Collaboration}

\address{
School of Physics and Nuclear Energy Engineering, Beihang University\\
Beijing, 100191, China\\
shencp@ihep.ac.cn}

\maketitle

%\begin{history}
%\received{Day Month Year}
%\revised{Day Month Year}
%\end{history}

\begin{abstract}
  My report consists of two parts: (1). Using samples of 102 million $\Upsilon(1S)$ and 158 million $\Upsilon(2S)$ events at Belle,
  we study 17 exclusive hadronic decays of these two bottomonium resonances to some Vector-Pseudoscalar (VP), Vector-Tensor (VT) and
  Axial-vector-Pseudoscalar (AP) processes and their final states. Branching fractions are measured for all the processes. The ratios
  of the branching fractions of $\Upsilon(2S)$ and $\Upsilon(1S)$ decays into the same final state are used to test a perturbative QCD (pQCD)
  prediction for OZI-suppressed bottomonium decays. (2). Using data samples of 89 fb$^{-1}$, 703 fb$^{-1}$, and 121 fb$^{-1}$ collected at
  center-of-mass (CMS) energies 10.52, 10.58, and 10.876~GeV, respectively, we measure the cross sections of
  $e^+e^- \to \omega\pi^0$, $K^{\ast}(892)\bar{K}$, and $K_2^{\ast}(1430)\bar{K}$.
  The energy dependence of the cross sections is presented.
\end{abstract}

\ccode{PACS numbers: 13.25.Gv, 14.40.Pq,  13.66.Bc, 13.40.Gp }

\section{Measurement of $\Upsilon(1S)$ and $\Upsilon(2S)$ decays
into VP final states}

We know for the OZI (Okubo-Zweig-Iizuka) suppressed
decays of the $J/\psi$ and $\psi(2S)$ to hadrons proceed via the
annihilation of the quark-antiquark pair into three gluons or a
photon, pQCD provides a relation for the ratios of branching fractions ($\cal B$)
for $\jpsi$ and $\psi(2S)$ decays
\begin{equation}
Q_{\psi}=\frac{{\cal B}_{\psi(2S) \to {\rm hadrons}}}{{\cal B}_{J/\psi \to {\rm hadrons}}}
=\frac{{\cal B}_{\psi(2S) \to e^+e^-}}{{\cal B}_{J/\psi \to
e^+e^-}} \approx 12\%,
\end{equation}
which is referred to as the ``12\% rule'' and is
expected to apply with reasonable accuracy to both inclusive
and exclusive decays. However, it is found to be severely violated for
$\rho\pi$ and other VP and VT final states.

A similar rule can be derived for OZI-suppressed bottomonium
decays:
 \begin{equation}
Q_{\Upsilon} =\frac{{\cal B}_{\Upsilon(2S) \to {\rm hadrons}}}{{\cal
B}_{\Upsilon(1S) \to {\rm hadrons}}} = \frac{{\cal B}_{\Upsilon(2S) \to
e^+e^-}}{{\cal B}_{\Upsilon(1S) \to e^+e^-}} = 0.77\pm 0.07.
 \end{equation}
%However, little experimental information is available on exclusive
%decays of the $\Upsilon$ resonances below $B\bar{B}$ threshold.

Recently, using 102 million $\Upsilon(1S)$ and 158 million
$\Upsilon(2S)$ events Belle studied exclusive hadronic decays of these two bottomonium resonances
to the three-body final states $\phi \kk$, $\omega \pp$ and
$K^{\ast 0}(892) K^- \pi^+ $~\cite{charge}, and to the two-body
VT states ($\phi f_2'(1525)$, $\omega f_2(1270)$, $\rho
a_2(1320)$ and $K^{\ast 0}(892) \bar{K}_2^{\ast 0}(1430) $) and
AP ($K_1(1270)^+ K^-$, $K_1(1400)^+
K^- $ and $b_1(1235)^+ \pi^- $) states~\cite{shen1}.
Signals are observed for the first time in the
$\Upsilon(1S) \to \phi K^+ K^-$, $\omega \pi^+ \pi^-$,
$K^{\ast 0} K^- \pi^+$, $K^{\ast0} K_2^{\ast 0}$ and
$\Upsilon(2S) \to \phi K^+ K^-$, $K^{\ast 0} K^- \pi^+$ decay modes.
Besides $K^{\ast0} K_2^{\ast 0}$,
no other two-body processes are observed in
all investigated final states.
For the
processes $\phi \kk$, $K^{\ast 0} K^- \pi^+ $, and $K^{\ast
0} \bar{K}^{\ast 0}_2(1430)$,  the $Q_{\Upsilon}$ ratios are consistent with
the expected value, while for $\omega \pp$, the measured $Q_{\Upsilon}$  ratio is
$2.6\sigma$ below the pQCD expectation. The results for the other modes
are inconclusive due to low statistical significance.

We also used the same data samples of $\Upsilon(1S)$ and $\Upsilon(2S)$ to study
exclusive hadronic decays to the
$\ks K^{+} \pi^-$, $\pp \pi^0 \pi^0$, and
$\pp \pi^0$, and two-body VP ($K^{\ast}(892)^0\bar{K}^0$,
$K^{\ast}(892)^-K^+$, $\omega\pi^0$, and $\rho\pi$) final states~\cite{shen2}.
After event selections, Fig.~\ref{vt-fit} shows the $K^+ \pi^-$
and $\ks \pi^-$ invariant mass distributions for the $\ks K^+ \pi^-$
final state, the $\pp \pi^0$ invariant mass distribution for
the $\pp\piz\piz$ final state, and
the $\pi\pi$ invariant mass
distribution for the $\pp\piz$ final state.  An unbinned simultaneous maximum likelihood fit
was done to these mass spectra. The results of the fits  are shown in
Fig.~\ref{vt-fit} and listed in Table~\ref{ta1}.

\begin{table}[ph]
\tbl{Results for the $\Upsilon(1S)$ and $\Upsilon(2S)$ decays,
where $N_{\rm sig}$ is the number of signal events from the fits, $N^{\rm
UL}_{\rm sig}$ is the upper limit on the number of signal events, $\Sigma$ is the statistical
significance ($\sigma$), $\BR$ is the branching fraction
(in units of $10^{-6}$),
$\BR^{\rm UL}$ is the 90\% C.L. upper limit on the branching fraction. }
{\begin{tabular}{c|ccc|ccc}
\hline
Channel & \multicolumn{3}{c|}{$\Upsilon$(1S)}
         &  \multicolumn{3}{c|}{$\Upsilon$(2S)}     \\
        & $N_{\rm sig}$/$N^{\rm UL}_{\rm sig}$ & $\Sigma$ & $\BR$/$\BR^{\rm UL}$  & $N^{\rm sig}$/$N^{\rm UL}_{\rm sig}$
        & $\Sigma$ &$\BR$/$\BR^{\rm UL}$
\\
\hline \rule{0mm}{0.4cm}
 $\ks K^+ \pi^-$ & $37.2\pm 7.6$&6.2&$1.59\pm0.33\pm0.18$
                 & $39.5\pm 10.3$&4.0&$1.14\pm0.30\pm 0.13$
                    \\
 $\pp \pi^0 \pi^0$&$143.2\pm22.4$&7.1&$12.8\pm2.01\pm2.27$
                  & $260.7\pm37.2$&7.4&$13.0\pm1.86\pm2.08$
                   \\
 $\pp \pi^0$ & $25.5\pm 8.6$   & 3.4 & $2.14\pm0.72\pm 0.34$
             & 15  & ---&  0.80
             \\\hline
 $K^{\ast0}\bar{K}^0$&$16.1\pm4.7$& 4.4&$2.92\pm0.85\pm0.37$
           &30&2.7& 4.22
          \\
 $K^{\ast-}K^+$&6.3&1.3&1.11
           &13&2.0&1.45
          \\
 $\omega \pi^0$&6.8& 1.6
               &3.90&4.6&0.1&1.63
               \\
 $\rho \pi$  & 22  & 2.2 &  3.68
        & 14   & --- & 1.16  \\
\hline
\end{tabular} \label{ta1}}
\end{table}

\begin{figure}[htbp]
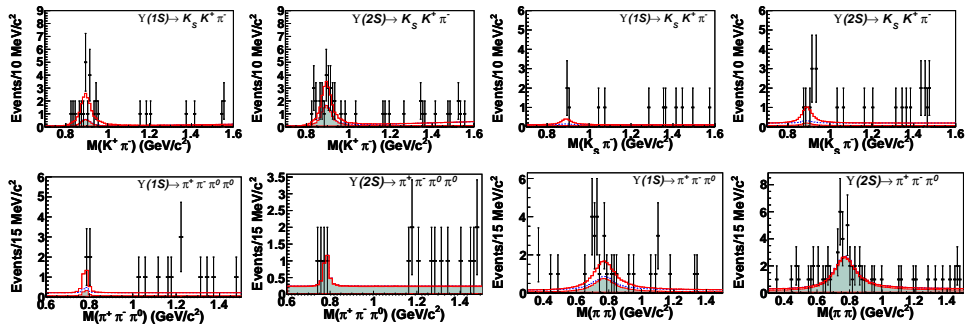

%\begin{figure*}
\includegraphics[height=3.1cm,angle=-90]{fig3a.epsi}
\includegraphics[height=3.1cm,angle=-90]{fig3b.epsi}
\includegraphics[height=3.1cm,angle=-90]{fig3c.epsi}
\includegraphics[height=3.1cm,angle=-90]{fig3d.epsi}\vspace{0.2cm}
\includegraphics[height=3.1cm,angle=-90]{fig3e.epsi}
\includegraphics[height=3.1cm,angle=-90]{fig3f.epsi}
\includegraphics[height=3.1cm,angle=-90]{fig3g.epsi}
\includegraphics[height=3.1cm,angle=-90]{fig3h.epsi}
\caption{\label{vt-fit} The fits to the $K^+ \pi^-$, $\ks \pi^-$,
$\pp \pi^0$ and $\pi \pi$ mass distributions for the $K^{\ast
}(892)^0$, $K^{\ast}(892)^-$, $\omega$ and $\rho$ vector meson
candidates from $\ks K^+ \pi^-$, $\pp \pi^0 \pi^0$ and $\pp \pi^0$
events from $\Upsilon(1S)$ and $\Upsilon(2S)$ decays.
The solid histograms show the results of the simultaneous fits, the
dotted curves show the total background estimates, and the shaded
histograms are the normalized continuum contributions.}
\end{figure}

Signals are observed for the first time in the
$\Upsilon(1S)\to \ks K^+ \pi^- $, $\pp \pi^0 \pi^0$ and
$\Upsilon(2S) \to \pp \pi^0 \pi^0$ decay modes.
There is an indication for large
isospin-violation between the branching fractions for the charged
and neutral $ K^{\ast}(892) {\bar K}$ for both $\Upsilon(1S)$ and
$\Upsilon(2S)$ decays, as in $\psp$ decays, which indicates
that the electromagnetic process plays an important role in these decays.
For the processes $\ks K^+ \pi^- $ and $\pi^+ \pi^- \pi^0 \pi^0$,
the $Q_{\Upsilon}$ ratios are consistent with the expected value;
for $\pp \pi^0$, the $Q_{\Upsilon}$ ratio is a little lower
than the pQCD prediction. The results for the other modes are
inconclusive due to low statistical significance.

\section{ Measurement of $e^+ e^- \to \omega \pi^0$, $K^{\ast}(892)\bar{K}$
and $K_2^{\ast}(1430)\bar{K}$}

Different models predict different energy dependence of the cross
sections for the process $\EE\to$ VP.
If SU(3) flavor symmetry is perfect, one expects the cross sections of $\omega\piz:K^{\ast
}(892)^0{\bar K}^0:K^{\ast}(892)^- K^+$  production equal 9:8:2. However, this relation was found to be
violated severely at $\sqrt{s}=3.67$ GeV and 3.773~GeV by the CLEO
experiment with the
ratio $R_{\rm VP}=\frac{\sigma_B(e^+e^-\to K^{\ast}(892)^0\bar
K^0)}{\sigma_B(e^+e^-\to K^{\ast}(892)^-K^+)}$ greater than 9
and 33 at $\sqrt{s}=3.67$ GeV and 3.773~GeV, respectively, at the 90\% C.L.
By taking into account  SU(3)$_{\rm f}$ symmetry breaking, a pQCD
calculation predicts $R_{\rm VP}=6.0$. In the quark model, one may
naively expect $R_{\rm TP}=\frac{\sigma_B(e^+e^-\to K_2^{\ast}(1430)^0\bar
K^0)}{\sigma_B(e^+e^-\to K_2^{\ast}(1430)^-K^+)}=R_{\rm VP}$.

The cross sections of $\EE\to \omega \pi^0$, $K^{\ast}(892) \bar{K}$, and
$K_2^{\ast}(1430) \bar{K}$ are measured~\cite{shen3}, based on data samples of 89~fb$^{-1}$, 703~fb$^{-1}$,
and 121~fb$^{-1}$ collected at $\sqrt{s}=$10.52, 10.58 ($\Upsilon(4S)$ peak), and
10.876~GeV ($\Upsilon(5S)$ peak), respectively. After event selections,  Fig.~\ref{vt-fit} shows the $\pp \pi^0$,
$K^+ \pi^-$, and $\ks \pi^-$ invariant mass distributions for the
$\pp\piz\piz$ and $\ks K^+ \pi^-$ final states. Unbinned maximum likelihood fitted
results are listed in Table~\ref{ta2} together with the calculated Born cross sections.
Assuming $1/s^{n}$ dependence, the fit gives $n=3.83\pm0.07$ and
$3.75\pm0.12$ for $\EE \to K^{\ast}(892)^0\bar{K}^0$
and $\omega \pi^0$ cross sections distributions.
With the calculated Born cross sections, we obtain $R_{\rm VP}>4.3, 20.0, 5.4,$ and
$R_{\rm TP}<1.1, 0.4, 0.6,$ for $\sqrt{s}=10.52$, 10.58, and 10.876~GeV, respectively, at the 90\% C.L.
Both the ratios are different from the predictions from exact or broken SU(3) symmetry models.

\begin{figure}[htbp]
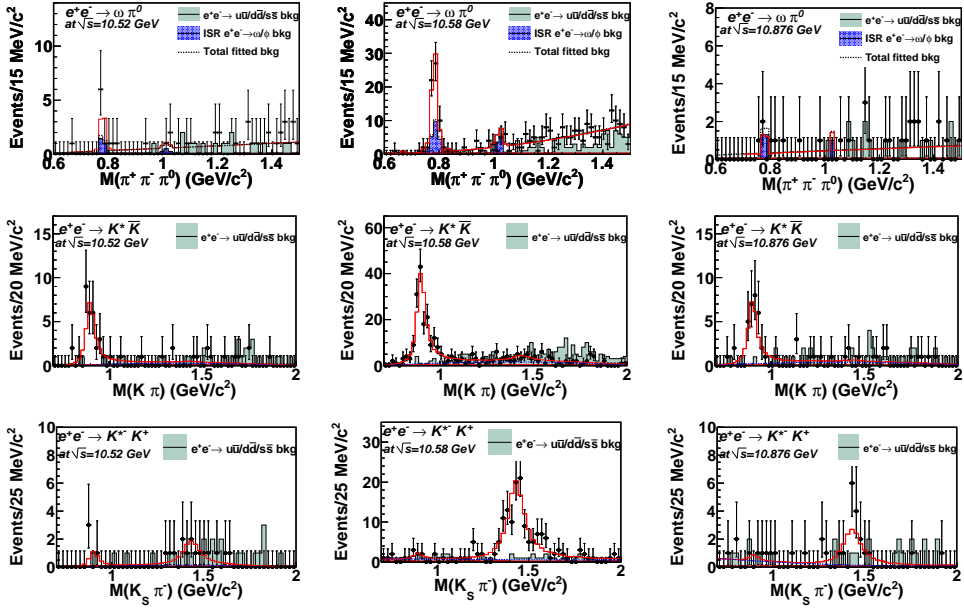

%\begin{figure*}
\includegraphics[height=3.9cm,angle=-90]{fig33a.epsi}
\includegraphics[height=3.9cm,angle=-90]{fig33b.epsi}
\includegraphics[height=3.9cm,angle=-90]{fig33c.epsi}\vspace{0.25cm}
\includegraphics[height=3.9cm,angle=-90]{fig33d.epsi}
\includegraphics[height=3.9cm,angle=-90]{fig33e.epsi}
\includegraphics[height=3.9cm,angle=-90]{fig33f.epsi}\vspace{0.25cm}
\includegraphics[height=3.9cm,angle=-90]{fig33g.epsi}\hspace{0.3cm}
\includegraphics[height=3.9cm,angle=-90]{fig33h.epsi}\hspace{0.45cm}
\includegraphics[height=3.9cm,angle=-90]{fig33i.epsi}
\caption{\label{vt-fit} The fits to the $\pp \pi^0$ (top row),
$K^+ \pi^-$ (middle row) and $\ks \pi^-$ (bottom row) invariant
mass distributions for the $\omega$, $K^{\ast}(892)$, and $K_2^{\ast}(1430)$
meson candidates from $\EE \to \pp \pi^0 \pi^0$ and $\ks K^+
\pi^-$ events from the $\sqrt{s}=10.52$ GeV, 10.58 GeV, and 10.876~GeV data
samples. The solid lines show the results of the
fits described in the text, the dotted curves show the total
background estimates, the dark shaded histograms are from the
normalized ISR backgrounds $\EE \to \gamma_{\rm ISR} \omega/\phi \to \gamma_{\rm ISR} \pp \pi^0$ and
the light shaded histograms are from the normalized $\EE \to
u\bar{u}/d\bar{d}/s\bar{s}$ backgrounds.}
\end{figure}

\begin{table}[ph]
\tbl{Results for the Born  cross sections, where $N_{\rm
sig}$ is the number of fitted signal events, $N^{\rm UL}_{\rm
sig}$ is the upper limit on the number of signal events, $\Sigma$ is the signal
significance, $\sigma_B$ is the Born  cross section,
$\sigma_B^{\rm UL}$ is the upper limit on the Born  cross
section. All the upper limits are given at the 90\% C.L. The
first uncertainty in $\sigma_B$ is statistical,
and the second systematic.}
{\begin{tabular}{c|cccccc}
\hline\hline
Channel & $\sqrt{s}$ (GeV) &$N_{\rm sig}$ & $N^{\rm UL}_{\rm sig}$& $\Sigma$ ($\sigma$) & $\sigma_B$ (fb)
        &$\sigma_B^{\rm UL}$ (fb)   \\ \hline
$\omega \pi^0$& 10.52 &$4.1^{+3.3}_{-2.6}$& 9.9& 1.6&$4.53^{+3.64}_{-2.88}\pm 0.50$ & 11 \\
              & 10.58 &$38.8^{+8.3}_{-7.6}$& ---&  6.7&$6.01^{+1.29}_{-1.18}\pm0.57$ &--- \\
              & 10.876 &$-0.7^{+2.9}_{-2.1}$& 7.0& ---&$-0.68^{+2.71}_{-1.97}\pm 0.20$ & 6.5 \\
\hline \rule{0mm}{0.4cm}
$K^{\ast}(892)^0\bar{K}^0$& 10.52 &$34.6^{+6.9}_{-6.1}$&---& 7.4&$10.77^{+2.15}_{-1.90}\pm0.77$&---\\
                    & 10.58 &$187\pm 17$&---& $>$10&$7.48\pm0.67\pm 0.51$&---\\
                    & 10.876&$34.6^{+7.5}_{-6.7}$&---& 7.2&$7.58^{+1.64}_{-1.47}\pm0.63$&---\\
\hline \rule{0mm}{0.4cm}
$K^{\ast}(892)^-K^+$& 10.52 & $4.6^{+3.6}_{-2.7}$ & 9.3  & 1.4 & $1.14^{+0.90}_{-0.67}\pm0.15$ & 2.3 \\
              & 10.58 & $5.9^{+4.7}_{-3.8}$ & 14  & 1.5 & $0.18^{+0.14}_{-0.12}\pm 0.02$ & 0.4\\
              & 10.876 & $1.6^{+3.9}_{-3.0}$ & 8.5  & 0.3 & $0.28^{+0.68}_{-0.52}\pm0.10$ & 1.5  \\
\hline \rule{0mm}{0.4cm}
$K_2^{\ast}(1430)^0\bar{K}^0$&10.52 &$1.3^{+4.3}_{-3.9}$ & 6.8  & 0.3 & $0.76^{+2.53}_{-2.26}\pm 0.14$ & 4.0 \\
                      &10.58 & $21^{+11}_{-10}$ & 40  & 2.1 & $1.65^{+0.86}_{-0.78}\pm 0.27$ & 3.1 \\
                      &10.876& $1.0^{+4.5}_{-3.7}$ & 8.9  & 0.2 & $0.38^{+1.79}_{-1.47}\pm0.07$ & 3.5   \\
\hline \rule{0mm}{0.4cm}
$K_2^{\ast}(1430)^-K^+$&10.52 & $12.0^{+6.2}_{-5.8}$ & 21  & 2.1 & $6.06^{+3.13}_{-2.93}\pm1.34$ & 11  \\
                      &10.58 & $129\pm 15$ & ---  & $>$10 & $8.36\pm0.95\pm0.62$ & --- \\
                      &10.876& $17.6^{+5.3}_{-4.6}$ &---  & 4.5 & $6.20^{+1.86}_{-1.63}\pm0.64$ & ---   \\
\hline\hline
\end{tabular} \label{ta2}}
\end{table}

%%%%%%%%%%%%%%%%%%%%%%%%%%%%%%%%%%%%%%%%%%%%%%%%%%%%%%%%%%%%%%%%%%%%%%%%%%%%%%%%%%%%%%%%%%%%
\section*{Acknowledgments}

This work is supported partly by the Fundamental Research Funds for the Central Universities of China (303236).

%%%%%%%%%%%%%%%%%%%%%%%%%%%%%%%%%%%%%%%%%%%%%%%%%%%%%%%%%%%%%%%%%%%%%%%%%%%%%%%%%%%%%
%\begin{thebibliography}{000} %for 3 digits
%\begin{thebibliography}{00}  %for 2 digits


\begin{thebibliography}{0}
\bibitem{charge} Charge-conjugate decays are implicitly assumed
throughout the report.
\end{thebibliography}
\end{document}